# Comprehensive assessment of error correction methods for high-throughput sequencing data

Yun Heo, Gowthami Manikandan, Anand Ramachandran, Deming Chen (dchen@illinois.edu)
University of Illinois at Urbana-Champaign

abstract
# ABSTRACT

The advent of DNA and RNA sequencing has revolutionized the study of genomics and molecular biology. Next generation sequencing (NGS) technologies like Illumina, Ion Torrent, SOLiD sequencing etc. have brought about a quick and cheap way to sequence genomes. Recently, third generation sequencing (TGS) technologies like PacBio and Oxford Nanopore Technology (ONT) have also been developed. Different technologies use different underlying methods for sequencing and are prone to different error rates. Though many tools exist for error correction of sequencing data from NGS and TGS methods, no standard method is available yet to evaluate the accuracy and effectiveness of these error-correction tools. In this study, we present a Software Package for Error Correction Tool Assessment on nuCLEic acid sequences (SPECTACLE) providing comprehensive algorithms to evaluate error-correction methods for DNA and RNA sequencing, for NGS and TGS platforms. We also present a compilation of sequencing datasets for Illumina, PacBio and ONT platforms that present challenging scenarios for error-correction tools. Using these datasets and SPECTACLE, we evaluate the performance of 23 different error-correction tools and present unique and helpful insights into their strengths and weaknesses. We hope that our methodology will standardize the evaluation of DNA and RNA error-correction tools in the future.

# INTRODUCTION

Rapid improvements in next-generation sequencing (NGS) technologies have allowed us to generate a huge amount of sequencing data at a low cost. However, the quality of the data has not improved at the same pace as the throughput of the NGS technologies. For example, one of the widely-used Illumina sequencing machines, HiSeq X Ten can produce 1.8 tera base pairs (bp) in each run, but only about 75 percent of the bases are guaranteed to have Phred scores of over 30 (http://www.illumina.com/systems/hiseq-x-sequencing-system/performance-specifications.ilmn).

Errors in NGS reads degrade the quality of downstream analyses, and correcting errors has been shown to improve the quality of these analyses [1-3]. Many standalone methods for correcting errors in DNA reads have been developed [4-19]. Besides, some DNA assemblers have their own error-correction modules, which can be used as standalone error-correction tools [20-22].

NGS is also applied to transcriptomic analysis [23]. RNA sequencing data also has sequencing errors, which makes RNA error correction an important problem to address. Error-correction methods for DNA reads may not work well for RNA sequencing data because of non-uniform expression levels and alternative splicing. To address this problem, Le, et al., [24] developed a new error-correction tool for RNA sequencing data.

Recently, several third-generation sequencing (TGS) technologies have been developed. TGS sequencing does not require amplification [25]. Single-molecule real-time (SMRT) sequencing technology from Pacific Biosciences and Oxford Nanopore (ONT) sequencing are widely used third-generation sequencing technologies. Even though sequencing systems that use the SMRT sequencing technology can generate reads up to tens of thousands of base pairs long, they have about 11% error rate and the errors are evenly distributed in reads [26]. Similarly, ONT's MinION reads have an error rate of 38.2% [27]. Also, the dominant error types of these technologies are insertions and deletions that are rare in Illumina reads. Due to these characteristics, dedicated error-correction methods for PacBio reads [28-31] and Oxford Nanopore reads [32-35] have been developed.

Despite such a large number of error-correction methods, only a few studies exist that are dedicated to the evaluation of the accuracy of these methods. Such scarcity is mainly due to the difficulty involved in discerning how many errors were corrected and how many were newly generated in the error-correction process. While checking if substitution errors have been corrected is easily done by measuring the Hamming distance between a reference sequence and a corrected read, it is not so simple to evaluate how accurately errors are corrected when insertions and deletions also exist as errors. The evaluation becomes more complex when reads are trimmed during error correction. Aligning a read to the source genome does not always provide the right solution since multiple best alignments can exist [5]. Heterozygosity also makes the evaluation hard. In a diploid genome, the same locus in a pair of chromosomes could have different alleles. Therefore, one of them can potentially be recognized as a sequencing error if reads from heterozygous regions are compared with one reference sequence.

To the best of our knowledge, only a handful of research works have been carried out to quantitatively evaluate how exactly errors in NGS reads have been corrected. Error Correction Evaluation Toolkit (ECET) [36] is an error-correction evaluation platform that consists of two software packages, one of which evaluates Illumina reads and the other, 454 or Ion Torrent reads. The reason for having two separate algorithms for dealing with different technologies is that the dominant error models of 454 and Ion Torrent reads are insertions or deletions in homopolymers while most errors in Illumina reads are substitutions [37, 38].

Another evaluation work by Molnar et al. [39] determines the correctness of reads or $k$-mers in the outputs from Illumina error-correction tools instead of directly checking the correctness of bases. Their method calculates (1) how many error-free reads or $k$-mers cover each base in a genome and (2) how many bases in a reference sequence are covered by error-free reads or $k$-mers, then checks how the two numbers are changed by error correction.

Another evaluation methodology, compute_gain, is a program that is a part of an error-correction tool package Fiona [5]. It aligns both a read and its corrected version to a reference sequence, and calculates the difference in edit distance between the two alignments. Ambiguities in alignments are resolved by placing gaps at the leftmost or rightmost possible position.

Even though the three methods opened up ways of evaluating the outputs from error-correction methods, all of them have limitations. The software package for Illumina reads in ECET can only work with the tools that explicitly specify the number of bases trimmed from both ends of reads. Even when this information is available, separate programs for each error-correction tool are needed to extract the number of trimmed bases, because the tools output the number in different ways.

The software package for 454 or Ion Torrent reads in ECET can evaluate reads with insertions and deletions but the evaluation results could be inaccurate for trimmed reads. Figure S2 in the supplementary document shows some examples in which ECET evaluates reads incorrectly.

Even though the software reported in [39] can be applied to the outputs from any Illumina error-correction method, it may not be applicable to other sequencing technologies. Since PacBio reads, for example, have a high error rate and the errors are evenly distributed in the reads, it is hard to get error-free *k*-mers of sufficient length. If short *k*-mers are used by this tool for the evaluation of PacBio reads, the specificity of the evaluation would be low because it is likely that the same or similar *k*-mers exist in other parts of the genome sequence as well.

The evaluation results of compute_gain, like that of ECET, could be inaccurate in some cases. Since the alignment scores used in compute_gain were designed to evaluate edit distance, a read could be aligned to a reference sequence in totally different ways before and after error correction, which makes it possible for the evaluation result to be inaccurate (see Figure S3 in the supplementary document).

In addition to these methods, evaluations are presented when individual error-correction tools are introduced. For Illumina tools, quality of error correction is evaluated using counts of uncorrected errors and corrected errors, and expressed in the form of metrics such as sensitivity, precision and gain [6-16]. The error counts used are obtained by mapping the reads to the reference, as reported in [36]. We outlined in paragraphs above how current mapping-based methods can give inaccurate evaluation results. Literature reporting new TGS error-correction tools on the other hand, do not report error count metrics such as sensitivity or gain. Instead, performance is measured based on improvements in downstream assembly and alignment results [28-25]. While such results give a good picture of how much improvement is made by error correction, there can be variations in such measurements based on the specific assembly or alignment tools used to obtain these metrics.

In this work, we try to address limitations of existing error-correction evaluation methods. We develop a new algorithm for evaluation of error-correction accuracy in terms of error count statistics that avoids the pitfalls of current alignment-based evaluation methods. This algorithm can be applied across different sequencing platforms from the NGS and TGS paradigms, allowing a uniform and standard error-correction evaluation method across technologies. In addition, we introduce new metrics for error-correction tool study that provide additional insights regarding design limitations of a given tool, giving pointers on how the tool may be improved. Using these new methods, we perform a comprehensive analysis of many state-of-the-art error correction tools and report error count statistics, alignment and assembly statistics, as well as additional metrics for understanding tool behavior. More specifically, the key contributions of this work can be summarized as follows:

1) We have developed a new error-correction tool evaluation algorithm that is generic across different sequencing error models and error rates. It can quantitatively evaluate any error-correction tool for NGS and TGS reads. It works for both DNA and RNA sequencing data, and can differentiate heterozygous alleles from sequencing errors.

2) We have designed input read sets that highlight the challenges in error correction such as heterozygosity, coverage variation, and repeats (the input datasets are available for download). These reads can be used as standard inputs for the evaluation of error-correction tools.

3) We have evaluated and compared 23 state-of-the-art error correction tools for NGS and TGS reads using SPECTACLE with these data sets as inputs. From SPECTACLE, we report many statistics pertaining to error correction like sensitivity, gain, precision, F1-Score, percentage similarity of reads, NG50 length, supporting read coverage, alignment quality of corrected reads, performance of the tool with respect to read position etc. for both NGS and TGS reads. This will give users systematic evaluations of strengths and weaknesses of the tools and indicate potential ways for their further improvement.

In the sections that follow, we will explain how we prepared the inputs for our evaluation and how the evaluation algorithm works. We then present and discuss the evaluation results and what should be done in the future.

## EVALUATION METHOD

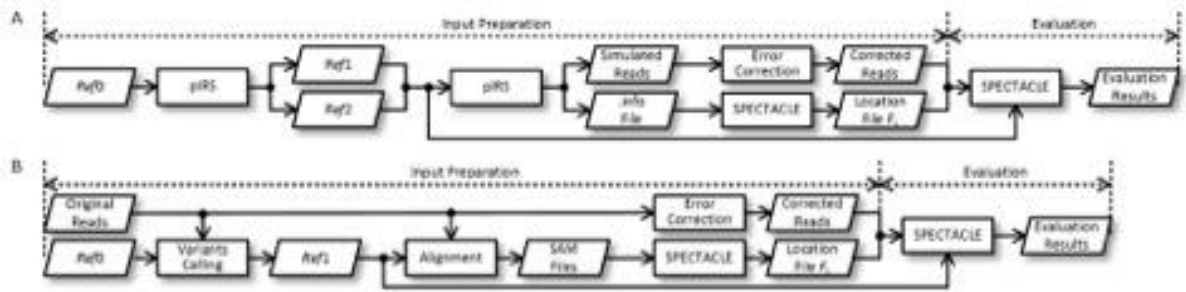

**Figure 1.** Flowchart of evaluating corrected DNA sequencing reads using SPECTACLE. (A) Evaluation flow for simulated reads. (B) Evaluation flow for real reads.

Figure 1 shows the SPECTACLE flows for evaluating error-correction tools with DNA simulated reads and DNA real reads. Each flow consists of two steps. In the first step, the locations of errors in input reads are determined, and in the next step this information is used to evaluate the output of an error-correction tool. The two steps will be explained in detail in the following subsections. The basic flow for evaluating RNA error-correction tools is similar and is explained in the supplementary document.

**Preparing Input Data**

SPECTACLE supports using both simulated reads and real reads to utilize their unique strengths. With simulated reads, we can determine the exact locations of errors in the reads. Moreover, reads can be generated from multiple reference sequences with some differences in order to check whether an error-correction tool is able to differentiate heterozygosity from sequencing errors.

The biggest advantage of using real reads is that no assumptions or modeling artifacts exist behind the sequencing data. Therefore, real reads can have some interesting properties that may not be accurately modeled in simulated reads. On the other hand, there can be ambiguities in finding error locations in real reads. In order to find the error locations in real reads, the reads need to be aligned to a reference sequence, and this can cause some problems. First, it is possible that a read can be aligned to multiple similar locations in a reference sequence (or to the same location in different ways), and determining the correct alignment is sometimes impossible. In the case of highly repetitive genomes, ambiguous alignments occur frequently, raising the chances of inaccurate evaluation results. Second, reads and a reference sequence might come from different samples, and the differences between them (i.e., variants) may be recognized as errors in this step. Third, the evaluation results will depend on the accuracy of the alignment tool.

SPECTACLE can work with the output reads from any read simulator that gives error location information in a Sequence Alignment/Map (SAM) format. However, in our study we used pIRS [40] exclusively for generating simulated Illumina reads. Error correction becomes challenging when there are heterozygosity and read coverage variations [3, 41], and pIRS can produce reads with these characteristics. pIRS can generate reads using a diploid genome, and the reads have both sequencing errors and heterozygosity. Second, pIRS can change read coverage depth of a specific genomic region according to the GC-content of the region.

Figure 1A depicts the evaluation flow for simulated reads. First, two reference sequences *Ref*1 and *Ref*2 that represent a pair of chromosomes in a diploid genome are generated by adding different variant sets to the input reference sequence *Ref*0. Once the two sequences are created, reads are generated from *Ref*1 and *Ref*2. The maximum ploidy level that SPECTACLE supports is two.

After the reads are generated, the locations of errors in the reads should be written in an error location file $F_L$. $F_L$ contains 1) the positions where reads originate in the genome, 2) the locations of substitutions, insertions, and deletions in each read, and 3) reference sequence from which each read was sampled (i.e. *Ref*1 or *Ref*2). When pIRS generates reads, it also produces a file containing the error locations (i.e. .info file) and .info file is converted into $F_L$.

In order to simulate PacBio reads, we used PBSIM [42]. PBSIM generates a Mutation Annotation Format (MAF) file for indicating error locations, and the file is converted to $F_L$. Because these TGS technologies do not use amplification techniques that cause higher error rates in regions in the genome that have higher G and C base content, the coverage variation due to different GC-content values was not considered in generating the simulated reads for PacBio. These TGS reads are generated from a single reference sequence because their error rate is much higher than the frequency of heterozygous sites and we do not expect the evaluation results to be altered appreciably by adding heterozygous points. However, a real dataset is used for evaluating PacBio reads that is heterozygous.

Figure 1B shows the evaluation flow for real reads. If input reads and a reference sequence *Ref*0 do not come from the same sample, there can be variants between them; the variants would be recognized later in the flow as sequencing errors. To overcome this problem, a new reference sequence, *Ref*1, is generated by calling the variants and applying them to *Ref*0. In our evaluation, BWA [43] and SAMtools [44] were used for variant calling. The variants are added to *Ref*0 using VCFtools [45], the input reads are aligned to *Ref*1, and the alignment results in the SAM file are converted to $F_L$. Among the substitution errors in $F_L$, the errors generated by heterozygous alleles are removed by comparing $F_L$ with the variant calling result.

**Evaluating the Accuracy of Corrected Reads**

Let $R_C$ be the corrected version of a read *R*. In order to evaluate the accuracy of $R_C$, we should find corrected errors and newly added errors in $R_C$. SPECTACLE first takes the segment $G_R$ from a reference sequence where read *R* was sampled. Then, $R_C$ is aligned to $G_R$ to find the errors in $R_C$ (errors missed, or introduced by a tool). For Illumina reads, we implemented a modified version of the Gotoh algorithm [46] for handling trimmed bases and for extracting all the alignments with the best alignment score. This algorithm is explained in detail in the supplementary document.

There can be a set of alignments, $ALN_{BEST}$, having the highest alignment score for a read $R_C$, but each alignment could imply different numbers of corrected and newly introduced errors. Instead of arbitrarily picking one of the alignments from $ALN_{BEST}$ to count the number of errors in $R_C$, SPECTACLE introduces additional criteria that rank each of the alignments in $ALN_{BEST}$ based on the error-correction accuracy in each case. Specifically, SPECTACLE calculates a penalty score based on newly introduced errors for each alignment in $ALN_{BEST}$, utilizing the scores used in the alignment step. Then, the alignment, *aln*<sub>BEST</sub>, from *ALN*<sub>BEST</sub> that has the least penalty is chosen. SPECTACLE makes the choice using the following equation, where *ERR*(*aln*) and *ERR*(*R*) are the sets of errors in an alignment *aln* and *R* and *ERR*(*aln*)\*ERR*(*R*) is the set of errors in *aln* but not in *R*.

$$aln_{BEST} = \underset{aln \in ALN_{BEST}}{\operatorname{argmax}} \sum_{err \in (ERR(aln) \setminus ERR(R))} penalty(err)$$

After *aln*<sub>BEST</sub> is chosen, we can compute from it, how many errors in *ERR*(*R*) are corrected and how many errors are newly added during correction.

In the case of Illumina reads, the evaluation of *aln*<sub>BEST</sub> from the set of best alignments, *ALN*<sub>BEST</sub>, follows a brute-force enumeration technique. However, for TGS reads that have long read lengths with indel errors and high error rates, there may be a very large number of candidates to be evaluated, making this enumeration method infeasible. To solve this problem for TGS reads, we used a different implementation that combines a dynamic programming-based enumeration of *aln*<sub>BEST</sub> with the alignment algorithm. However, to simplify the implementation it currently uses a fixed and simpler scoring scheme, namely, a score of -1 is assigned to gaps and mismatches and a score of +1 is assigned to matches. Using this implementation, we can evaluate the same error count statistics for TGS reads as for Illumina reads, albeit with a limited scoring option. We plan to improve this to provide scoring flexibility in the future. The dynamic programming recursions are explained in detail in the supplementary document.

In order to classify the bases in input reads, we introduce a notation consisting of a triplet, each character of which should be either Y or N. The first character indicates whether the base in the original read is correct (Y) or not (N), the second character indicates whether the base has been modified by an error correction tool (Y) or not (N), and the third one indicates whether the base in the corrected read at that position is correct (Y) or not (N). For example, NYY describes a base that is erroneous in $R$, modified by an error correction tool, and error-free in $R_C$. All the bases should fall into one of the five categories: NNN, NYN, NYY, YNY, and YYN because YYY, YNN, and NNY are logically meaningless. Using these triplets, the accuracy metrics that are summarized in Table 1 are calculated.

**Table 1.** Accuracy metrics.

| Metrics | Equations |
|---|---|
| Sensitivity | sum(NYY) / (sum(NYY) + sum(NYN) + sum(NNN)) |
| Gain | (sum(NYY) - sum(YYN) - sum(NYN)) / (sum(NYY) + sum(NYN) + sum(NNN)) |
| Specificity | sum(YNY) / (sum(YYN) + sum(YNY)) |
| Precision | sum(NYY) / (sum(NYY) + sum(YYN) + sum(NYN)) |
| F-score | 2 sum (NYY) / (sum(NYY) + sum(YYN) + 2sum(NYN) + sum(NNN)) |

SPECTACLE also can calculate and report the percentage similarity of reads for error correction evaluation. This feature is mainly intended for long reads. Percentage similarity of a read set $S_R$ is defined using the following equation, where $N_{RM}$, $N_{RMM}$, $N_{RI}$, and $N_{RD}$ are the number of matched bases, the number of mismatched bases, the number of inserted bases, and the number of deleted bases in the alignment result of $R$, respectively:

$$\text{Percentage Similarity} = \sum_{R \in S_R} \frac{N_{RM}}{N_{RM}+N_{RMM}+N_{RI}+N_{RD}}$$

SPECTACLE calculates percentage similarity both for input reads and for their error correction results, and shows how this number is improved after error correction.

Most TGS error correction methods trim uncorrected regions in reads. After this process, $R_C$ could be split into multiple pieces and become much shorter than $R$. Therefore, SPECTACLE also reports read coverage that indicates how long total read length (after trimming) is and NG50 [1] that shows how long the average read length is.

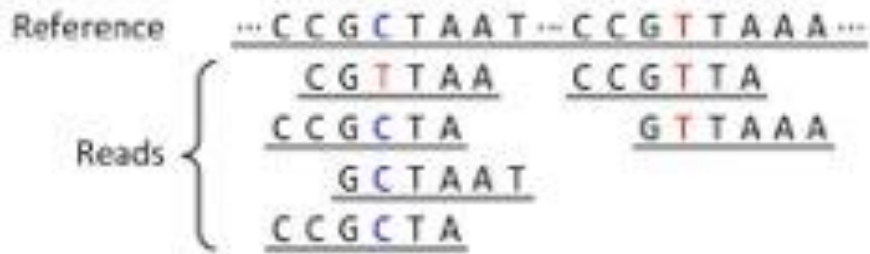

**Figure 2.** Supporting reads and supporting read coverage. Supporting reads are the reads that include a specific position of a reference sequence with a specific base at the position. In the left side, there is a read CGTTAA with an erroneous base T, and three more correct reads are also sampled there. In this example, the number of supporting reads (i.e. supporting read coverage) for T at that position of the reference sequence is 1, while supporting read coverage for C is 3. However, there is another similar sequence in the reference sequence (i.e. repeats) and the reads sampled at the right region could be supporting read for T at the left side, which makes it hard to correct the error. Differential supporting read coverage of an erroneous base can be defined as (supporting read coverage of correct base) - (supporting read coverage for the erroneous base).

In addition to these metrics, SPECTACLE can report other detailed analyses such as those related to supporting read coverage which help users understand the characteristics of an error-correction tool in depth. Figure 2 explains a supporting read, supporting read coverage, and differential supporting read coverage. An error in a read becomes difficult to correct if the corresponding correct base has low supporting read coverage. This is because most error-correction tools recognize bases with low supporting read coverage as errors. Low differential supporting read coverage also makes error correction harder, because then both a correct base and an erroneous base have a similar number of supporting reads. SPECTACLE gives the percentage of corrected bases against supporting read coverage for correct bases, and the percentage of corrected bases against differential supporting read coverage. This metric helps in evaluating how sensitive an error correction tool is to variations in read coverage.

SPECTACLE collects the percentage of corrected bases in each position of reads (i.e., point sensitivity). Based on this, users can judge whether an error-correction tool can correct errors in a specific region of reads or not. This report can allow SPECTACLE users to discern how the output of an error-correction tool can be polished further, how multiple error-correction algorithms can be combined, and how an error-correction algorithm can be improved.

SPECTACLE also reports measurements that provide an idea about how good the corrected reads are in the context of downstream analyses. One of the most intuitive ways to evaluate these is to count the number of corrected reads that can be aligned to a reference sequence without mismatches or indels. However, this result can be misleading when reads are aligned to wrong positions in a reference sequence. In order to avoid this, SPECTACLE has the capability to compare the aligned locations of reads in a SAM format with $F_L$. If insertions or deletions in a read are corrected, the aligned position of the read can be shifted. SPECTACLE determines the largest possible amount of shift in the aligned positions for each read using the number of insertions and deletions, and then reports the number of reads aligned correctly within this predicted range.

The average number of times each base in the reference sequence is covered by error-free reads (i.e. error-free read coverage) and the fraction of a reference sequence that is covered by error-free reads (i.e. chromosome coverage) are important metrics that indicate the quality of a read set [39]. SPECTACLE collects the two numbers using the exact alignment result described above.

# RESULTS

We evaluated 17 Illumina read error-correction tools, four PacBio and two ONT read error-correction methods using SPECTACLE. All the experiments were done on a cluster, each computing node of which had two six-core Intel Xeon X5650 processors and 24 GB of memory.

In the following sections, we have included only selected results that highlight the strengths and weaknesses of the tools. The remaining results, software versions, and software command line options are available in the supplementary document.

## DATA PREPARATION

**Preparing Illumina Read Sets**

As discussed above, coverage variation, heterozygosity, and repeats complicate error correction, and all the three factors were considered when we prepared input reads for our evaluation. The Illumina read sets we prepared are described in Table 2. Five different genomes I1-I5 were used to generate simulated read sets. Even though high coverage read sets are popular, correcting errors in low-coverage reads is still important. For example, cancer genome samples could be the mixture of cancer genomes and normal genomes, and the portion of one of the genomes could be very low [47]. Error-correction tools for such genomes should have the capability to correct errors in low-coverage reads. Therefore, read sets having both high and low coverage values are considered, and the coverage value is indicated using the postfixes -10X, -20X, -30X, and -40X. Coverage ranges from 10x to 40x have been picked to be consistent with base datasets used in other works reporting and validating error-correction methods.

Table 2. Details of Illumina read sets.

| ID | Reference | | | | | Read | | |
|---|---|---|---|---|---|---|---|---|
| | Species | Accession Number | $G_L$ (Mbp) | GC (%) | | Length | Cov. (X) | Error Rate (%) |
| | | | | Avg. | Std. | | | |
| I1-10X | *R. sphaeroids* | NC_007488.1 NC_007489.1 NC_007490.1 NC_007493.1 | 4.6 | 68.8 | 6.3 | 100 | 10 | 0.4 |
| I1-20X | | | | | | 100 | 20 | 0.4 |
| I1-30X | | | | | | 100 | 30 | 0.4 |
| I1-40X | | | | | | 100 | 40 | 0.4 |
| I2-10X | *B. cereus* ATCC 10987 | NC_003909.8 NC_005707.1 | 5.4 | 35.5 | 6.3 | 100 | 10 | 0.4 |
| I2-20X | | | | | | 100 | 20 | 0.4 |
| I2-30X | | | | | | 100 | 30 | 0.4 |
| I2-40X | | | | | | 100 | 40 | 0.4 |
| I3-10X | *O. sativa Chr. 5* | NC_008398.2 | 29.9 | 44.0 | 13.5 | 100 | 10 | 0.4 |
| I3-20X | | | | | | 100 | 20 | 0.4 |
| I3-30X | | | | | | 100 | 30 | 0.4 |
| I3-40X | | | | | | 100 | 40 | 0.4 |
| I4-10X | *Mouse Chr. Y* | NC_000087.7 | 88.1 | 38.9 | 8.0 | 100 | 10 | 0.4 |
| I4-20X | | | | | | 100 | 20 | 0.4 |
| I4-30X | | | | | | 100 | 30 | 0.4 |
| I4-40X | | | | | | 100 | 40 | 0.4 |
| I5-10X | *Human Chr. 1* | NC_00001.11 | 230.5 | 41.7 | 10.6 | 100 | 10 | 0.4 |
| I5-20X | | | | | | 100 | 20 | 0.4 |
| I5-30X | | | | | | 100 | 30 | 0.4 |
| I5-40X | | | | | | 100 | 40 | 0.4 |
| I6 | *B. cereus* ATCC 10987 | NC_003909.8 NC_005707.1 | 5.4 | 35.5 | 6.3 | 100 | 40 | 0.2 |

[$G_L$] genome length without Ns; [GC Avg.] average GC contents; [GC Std.] GC contents standard deviation; [Cov.] read coverage; [Error Rate] (<total number of substitutions> + <total number of inserted bases> + <total number of deleted bases>) / <total number of bases in reads>.

I1, I2, and I3 are *E. Coli* bacterium genomes that have different GC-content values. I4 is the mouse chromosome Y known as a highly repetitive genome [48]. I5 is human chromosome 1, the largest genome sequence used in our experiments.

To evaluate the results for real reads, we downloaded I6 from the Illumina website (http://www.illumina.com/systems/miseq/scientific_data.ilmn). The reads from this dataset have been sequenced from the exact same strain as I2 using the Illumina MiSeq sequencer.

Because the coverage of the reads is over 2,500 X, we down-sampled the reads to 40 X. Details regarding the down-sampling can be found in the supplementary document.

**Preparing PacBio Read Sets**

The read sets used for evaluating PacBio error-correction tools are shown in Table 3. The PacBio error-correction tools evaluated in this study require, in addition to PacBio reads, Illumina reads that are much more accurate than the PacBio reads as most PacBio error-correction tools use high quality Illumina short reads to detect and correct errors. These Illumina reads are described in the "Illumina" column of Table 3. In order to evaluate the effect of Illumina read coverage on the accuracy of error correction for PacBio reads, we prepared four different Illumina read sets with different read coverage values (suffixed -10X, -20X, -30X, and -40X). 40X-EF is an error-free version of the 40X read set and was used to evaluate the effects of sequencing errors in Illumina reads on error correction for PacBio reads.

**Table 3.** Details of PacBio read sets.

| ID | Reference | | | PacBio | | | Illumina | | | |
|---|---|---|---|---|---|---|---|---|---|---|
| | Species | Accession Number | $G_L$ (Mbp) | Length (bp) | Cov. (X) | Error Rate (%) | Accession Number | Length (bp) | Cov. (X) | Error Rate (%) |
| P1-10X | E. coli | NC_000913.3 | 4.6 | 500-14,494 | 21 | 23.4 | SRR922409 | 97 | 10 | 0.7 |
| P1-20X | | | | | | | SRR922409 | 97 | 20 | 0.6 |
| P1-30X | | | | | | | SRR922409 | 97 | 30 | 0.6 |
| P1-40X | | | | | | | SRR922409 | 97 | 40 | 0.6 |
| P2-10X | Human Chr19 10 Mbp | NC_000019.10 | 10.0 | 500-15,000 | 20 | 20.3 | N/A | 100 | 10 | 0.4 |
| P2-20X | | | | | | | N/A | 100 | 20 | 0.4 |
| P2-30X | | | | | | | N/A | 100 | 30 | 0.4 |
| P2-40X | | | | | | | N/A | 100 | 40 | 0.4 |
| P2-40X-EF | | | | | | | N/A | 100 | 40 | 0.0 |

[$G_L$] genome length; [Cov.] read coverage; [Error Rate] (<total number of substitutions> + <total number of inserted bases> + <total number of deleted bases>) / <total number of bases in reads>.

P1 is *E. coli* K12 M1665 strain, and both the PacBio reads and the Illumina reads are real reads. The PacBio reads were downloaded from Pacific Biosciences DevNet (https://github.com/PacificBiosciences/DevNet/wiki/E%20coli%20K12%20MG1655%20Hybrid%

20Assembly). Four Illumina read sets with different read coverage values were generated by taking different number of reads from SRR922409.

P2 is the first 10 Mbp region of human chromosome 19, which was used for evaluating the scalability of the PacBio error correction tools. We first tried using the entire human chromosome 19. However, only LoRDEC could be finished within 70 hours, which is the maximum allocated run time in our cluster; as a result, we had to use a portion of the chromosome. The PacBio reads and the Illumina reads for P2 were simulated using PBSIM and pIRS, respectively.

PacBio reads with lengths shorter than 500 bp were filtered out because short PacBio reads have a lot of errors (due to the high error rates in TGS sequencing) compared to the corresponding Illumina reads that have similar lengths. Hence, short PacBio reads would not have any major benefit compared to paired-end Illumina reads and are ignored for correction and evaluation.

**Preparing Oxford Nanopore (ONT) Read Sets**

Table 4 shows the details of the ONT datasets that have been used for the evaluation of ONT error-correction tools. ONT is a relatively newer technology, and ONT read simulation and error-correction techniques are currently maturing. As a result, are fewer ONT datasets that are publicly available for testing and evaluation. ONT error-correction tools also use short Illumina reads of high quality for error correction and the details of the Illumina reads used have been mentioned in Table 4. Both O1 and O2 are real read sets. O1 is *E. Coli K12 M1665* strain. The raw reads were downloaded from GigaDB (http://gigadb.org/dataset/view/id/100102/token/S30Hp9ZurcARyhov). O2 is *Saccharomyces* cerevisiae W303 strain downloaded from the NCBI Sequence Read Archive (http://www.ncbi.nlm.nih.gov/sra). The real Illumina reads for both these datasets were downloaded from Illumina BaseSpace (SRR567755). Similar to PacBio reads, ONT reads shorter than 500 bp were filtered out.

Table 4. Details of ONT (MinION) read sets.

| ID | Reference | | | MinION reads | | Illumina | | | |
|---|---|---|---|---|---|---|---|---|---|
| | Species | Accession Number | $G_L$ (Mbp) | Length (bp) | Error Rate (%) | Accession Number | Length (bp) | Cov. (X) | Error Rate (%) |
| O1-10X | E. coli | NC_000913.3 | 4.6 | 500-47,422 | 30.4 | SRR922409 | 97 | 10 | 0.7 |
| O1-20X | | | | | | SRR922409 | 97 | 20 | 0.6 |
| O1-30X | | | | | | SRR922409 | 97 | 30 | 0.6 |
| O1-30x-EF | | | | | | SRR922409 | 97 | 30 | 0.6 |
| O2-10X | Saccharomyces cerevisiae W303 | SRP055987 | 7.5 | 500-191,145 | 36.2 | SRR567755 | 250 | 10 | 0.02 |
| O2-20X | | | | | | SRR567755 | 250 | 20 | 0.02 |
| O2-30X | | | | | | SRR567755 | 250 | 30 | 0.02 |
| O2-30X-EF | | | | | | SRR567755 | 250 | 30 | 0.0 |

[$G_L$] genome length; [Error Rate] (<total number of substitutions> + <total number of inserted bases> + <total number of deleted bases>) / <total number of bases in reads>.

## RUNNING ERROR CORRECTION TOOLS

**Running Illumina Read Error-Correction Tools**

The input read sets were corrected using the 17 error-correction tools that had shown good accuracy in the previous evaluations or had been newly published at the time of running the evaluations. Among these, the stand-alone error correction tools are BFC [49], BLESS [6], Blue [7], Coral [8], HiTEC [9], Fiona[5], Lighter [10], Musket [11], Quake [12], QuorUM [13], RACER [14], Reptile [15],Trowel [16] and ECHO [17]. The remaining three tools are parts of DNA assemblers, ALLPATHS-LG [20], SGA [21], and SOAPdenovo [22].

For each error-correction method, we applied successive numbers to the key parameters of the tools, and generated multiple corrected output read sets corresponding to each parameter. The output read sets were assessed using SPECTACLE and the one that had the highest gain for substitutions, insertions, and deletions was chosen. The maximum *k*-mer length for Quake was limited to 18 beyond which the memory capacity of our server was exhausted.

ALLPATHS-LG, BFC, BLESS, Blue, Musket, Quake, QuorUM, RACER, Reptile, SGA, and SOAPec succeeded in generating outputs for all the input read sets. Coral, HiTEC, Fiona, and Trowel failed to correct errors in large genomes because of insufficient memory. ECHO had not finished after 70 hours for the I4 and I5 read sets. Lighter finished correcting all the read sets but it made no correction for the read sets with 10 X coverage.

Even though SPECTACLE can assess the outputs from error correction tools for RNA sequencing reads, the evaluation results for such tools have been excluded from the main document. SEECER [24] is the only software available for correcting RNA sequencing reads; for some input parameters, however, SEECER would occasionally terminate abnormally. So we ran SPECTACLE with SEECER with the parameters for which the tool had completed execution, and did not perform a parameter-space exploration. The results are in the supplementary document.

**Running TGS Read Error-Correction Tools**

Widely used PacBio read error correction tools LoRDEC [28], LSC [29], PBcR [30], and Proovread [31] were evaluated using P1 and P2. No parameter tuning was needed for LSC, PBcR, and Proovread. For LoRDEC, we generated multiple output sets by applying successive values for *k*-mer length and solid *k*-mer occurrence threshold, and chose the result that gave the highest percentage similarity explained in the method section. We could not assess LSC using P2 because it had not finished after 70 hours.

Since ONT is a relatively new technology, ONT read error correction technologies are just being explored and studied in detail. We evaluated two of the most recent ONT read error correction technologies NanoCorr [33] and NaS [32] using O1 and O2. Default parameters were used for each of the error-correction methods.

**EVALUATION RESULTS FOR ILLUMINA ERROR CORRECTION TOOLS**

**Accuracy of Illumina Error Correction Tools**

Table 5. Sensitivity and gain of substitution errors for the 40 X Illumina read sets.

| Software | I1-40X | | I2-40X | | I3-40X | | I4-40X | | I5-40X | | I6 | |
|---|---|---|---|---|---|---|---|---|---|---|---|---|
| | Sens. | Gain | Sens. | Gain | Sens. | Gain | Sens. | Gain | Sens. | Gain | Sens. | Gain |
| ALLPATHS-LG | 0.998 | 0.983 | 0.998 | 0.984 | **0.990** | 0.966 | 0.851 | 0.690 | 0.969 | 0.904 | 0.960 | 0.958 |
| BFC | 0.964 | 0.964 | 0.960 | 0.959 | 0.948 | 0.940 | 0.777 | 0.711 | 0.934 | 0.920 | 0.981 | **0.979** |
| BLESS | 0.998 | **0.997** | 0.998 | **0.998** | **0.990** | **0.983** | **0.905** | **0.855** | **0.975** | **0.964** | 0.979 | 0.977 |
| Blue | 0.998 | 0.961 | 0.998 | 0.970 | 0.981 | 0.883 | 0.850 | 0.520 | 0.896 | 0.819 | **0.982** | 0.903 |
| Coral | 0.979 | 0.913 | 0.987 | 0.934 | N/A | N/A | N/A | N/A | N/A | N/A | 0.817 | 0.806 |
| ECHO | 0.831 | 0.784 | 0.949 | 0.900 | 0.856 | 0.803 | N/A | N/A | N/A | N/A | 0.831 | 0.822 |
| Fiona | 0.998 | 0.973 | 0.998 | 0.980 | 0.984 | 0.902 | 0.677 | 0.237 | N/A | N/A | 0.970 | 0.967 |
| HiTEC | 0.997 | 0.982 | 0.997 | 0.993 | N/A | N/A | N/A | N/A | N/A | N/A | 0.965 | 0.959 |
| Lighter | 0.995 | 0.992 | 0.996 | 0.995 | 0.974 | 0.966 | 0.656 | 0.586 | 0.939 | 0.913 | 0.973 | 0.971 |
| Musket | 0.996 | 0.995 | 0.996 | 0.995 | 0.973 | 0.964 | 0.773 | 0.698 | 0.909 | 0.886 | 0.958 | 0.955 |
| Quake | 0.988 | 0.988 | 0.990 | 0.990 | 0.973 | 0.970 | 0.856 | 0.830 | 0.920 | 0.913 | 0.738 | 0.736 |
| QuorUM | **0.999** | **0.997** | **0.999** | **0.998** | 0.981 | 0.969 | 0.779 | 0.709 | 0.951 | 0.925 | **0.982** | 0.977 |
| RACER | 0.996 | 0.913 | 0.997 | 0.968 | 0.961 | 0.708 | 0.587 | -0.097 | 0.902 | 0.114 | 0.967 | 0.946 |
| Reptile | 0.958 | 0.933 | 0.968 | 0.960 | 0.926 | 0.824 | 0.672 | 0.562 | 0.878 | 0.760 | 0.852 | 0.831 |
| SGA | 0.996 | 0.996 | 0.996 | 0.996 | 0.975 | 0.968 | 0.738 | 0.673 | 0.959 | 0.939 | 0.947 | 0.944 |
| SOAPec | 0.671 | 0.670 | 0.664 | 0.664 | 0.650 | 0.648 | 0.478 | 0.446 | 0.624 | 0.614 | 0.539 | 0.538 |
| Trowel | 0.817 | 0.814 | 0.836 | 0.833 | 0.835 | 0.818 | 0.599 | 0.469 | N/A | N/A | 0.677 | 0.675 |

[Sens.] sensitivity.

Sensitivity and gain for substitution errors for the 40 X input read sets are summarized in Table 5. For all the bacterium genomes I1, I2, and I3, ALLPATHS-LG, BLESS, Lighter, Musket, Quake, QuorUM, and SGA generated outputs with gain above 0.95. For the highly repetitive genome I4, BLESS and Quake outperformed the others, and only these two tools obtained gain above 0.8.

For I5, the largest input genome, ALLPATHS-LG, BFC, BLESS, Lighter, Musket, Quake, QuorUM, and SGA showed gain above 0.9. Other than BFC, these are the same tools that worked well for I1-I3. In the evaluation using I6, most tools showed similar performance as they did for I2 since both I2 and I6 were generated from *B. cereus*. However, Coral, Quake, Reptile, SOAPec, and Trowel showed a degradation of above 0.1 for the gain value in I6 when compared with I2.

The difference between sensitivity and gain shows how many false corrections were made by each tool. In general, BFC, BLESS, Quake, SGA, and SOAPec generated fewer false corrections than the others.

Table 6. Sensitivity and gain of substitution errors for the I5 read sets with different coverage values.

| Software | I5-10X | | I5-20X | | I5-30X | | I5-40X | |
|---|---|---|---|---|---|---|---|---|
| | Sensitivity | Gain | Sensitivity | Gain | Sensitivity | Gain | Sensitivity | Gain |
| ALLPATHS-LG | 0.911 | 0.811 | **0.964** | 0.886 | 0.968 | 0.897 | 0.969 | 0.904 |
| BFC | 0.810 | 0.749 | 0.919 | 0.891 | 0.929 | 0.912 | 0.934 | 0.920 |
| BLESS | 0.931 | **0.898** | 0.961 | **0.946** | **0.975** | **0.960** | **0.975** | **0.964** |
| Blue | 0.848 | 0.690 | 0.894 | 0.809 | 0.896 | 0.818 | 0.896 | 0.819 |
| Fiona | **0.942** | 0.837 | N/A | N/A | N/A | N/A | N/A | N/A |
| Lighter | N/A | N/A | 0.918 | 0.867 | 0.938 | 0.907 | 0.939 | 0.913 |
| Musket | 0.889 | 0.860 | 0.905 | 0.882 | 0.907 | 0.885 | 0.909 | 0.886 |
| Quake | 0.908 | 0.896 | 0.917 | 0.910 | 0.920 | 0.912 | 0.920 | 0.913 |
| QuorUM | 0.894 | 0.810 | 0.952 | 0.907 | 0.952 | 0.922 | 0.951 | 0.925 |
| RACER | 0.819 | -2.287 | 0.898 | -0.164 | 0.902 | 0.052 | 0.902 | 0.114 |
| Reptile | 0.805 | 0.612 | 0.869 | 0.728 | 0.876 | 0.754 | 0.878 | 0.760 |
| SGA | 0.852 | 0.803 | 0.941 | 0.917 | 0.955 | 0.936 | 0.959 | 0.939 |
| SOAPec | 0.585 | 0.545 | 0.622 | 0.609 | 0.624 | 0.613 | 0.624 | 0.614 |

Table 6 shows the variation in gain with different read coverage values for I5. Only BLESS, Musket, and Quake generated gain above 0.85 for all the read sets. Lighter showed good results for 20-40 X reads, but it could not correct the errors in I5-10X. BFC, BLESS, Musket, Quake, SGA, and SOAPec made a small number of false corrections for low coverage read sets. Gain was saturated in most tools when read coverage became 30 X.

**Figure 3.** The percentage of corrected errors in I5-40X for various supporting read coverage of correct bases.

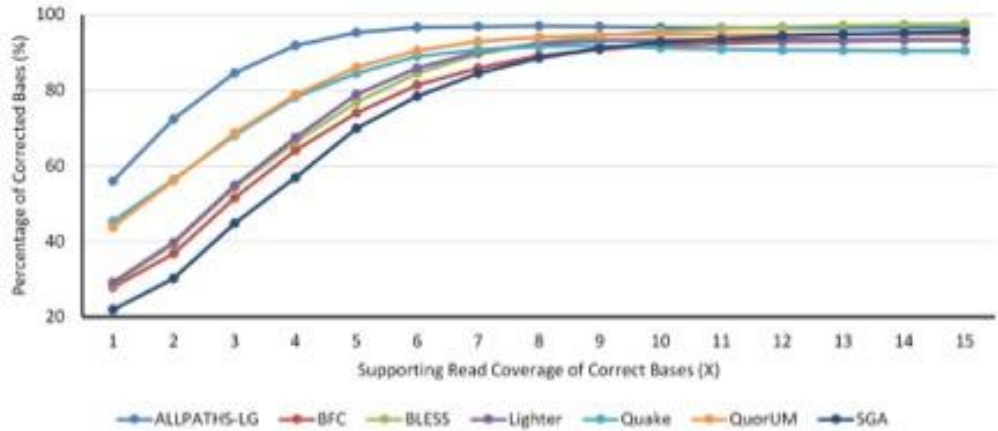

The percentage of corrected bases as a function of supporting read coverage for I5-40X is shown in Figure 3. ALLPATHS-LG, Quake, and QuorUM corrected more errors than the others when supporting read coverage of correct bases was close to 1. Even though ALLPATHS-LG and QuorUM have the capability to correct errors with low supporting read coverage, gain for I5-10X of the tools in Table 6 was not as impressive as this result. This is because they also generated many false positives for this input set.

The effect of differential supporting read coverage on error correction was significant only when read coverage was low. These results can be found in Figure S4 of the supplementary document.

As shown in Figure 4, tools can correct different percentages of errors in different locations in reads. The plots for ALLPATHS-LG, BFC, BLESS, and Lighter show relatively flat lines, which means that they corrected almost the same proportion of errors in all the positions in reads. On the other hand, plots for QuorUM and SGA have deep valley points, and the positions of these regions with little correction match with the *k*-mer length used with these tools for generating the respective outputs. In addition, Quake could only correct a relatively small number of errors at both ends of reads compared to the others.

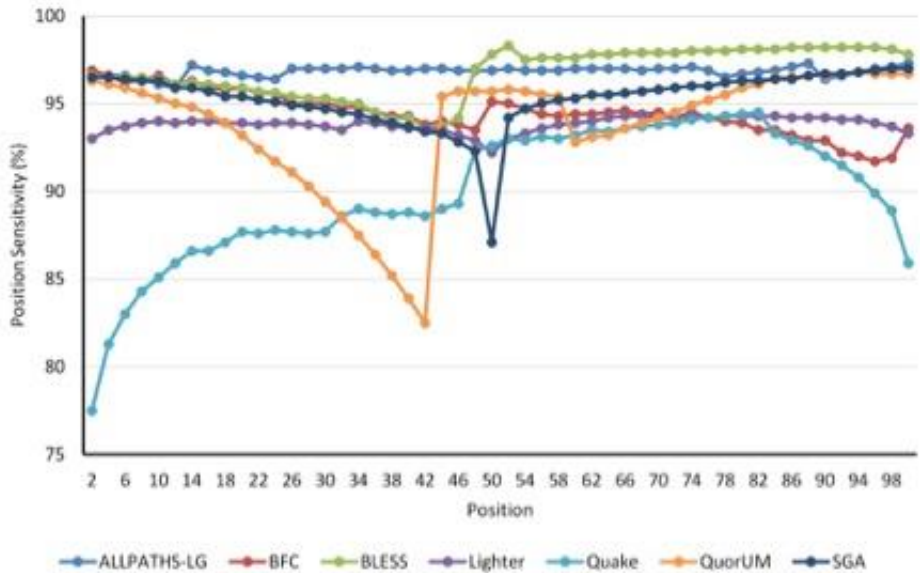

**Figure 4.** Point sensitivity of the I5-40X reads.

We also analyzed gain with respect to insertions and deletions and the results can be found in the supplementary document.

**Alignment Results for Illumina Error Correction Tools**

Table 7 shows how many corrected reads can be exactly aligned to the reference sequences. Reads were aligned using the paired-end alignment feature of Bowtie [50] without allowing any mismatches or indels. The genomes I1-I5 have two reference sequences, and corrected read sets were aligned to the reference sequence from which they originated. The alignment results are well matched with the results in Table 5, and the tools that showed high sensitivity also had more reads aligned correctly to the reference sequences.

**Table 7.** Alignment results of 40 X Illumina read sets.

| Software | I1-40X | | I1-40X | | I3-40X | | I4-40X | | I5-40X | | I6 | |
|---|---|---|---|---|---|---|---|---|---|---|---|---|
| | Aligned | Correct | Aligned | Correct | Aligned | Correct | Aligned | Correct | Aligned | Correct | Aligned | Correct |
| Original | 52.52 | 100.00 | 50.86 | 100.00 | 51.16 | 99.99 | 51.26 | 99.54 | 51.12 | 99.98 | 81.07 | 100.00 |
| ALLPATHS-LG | 99.07 | 99.98 | 99.07 | 99.97 | 98.51 | 99.93 | 88.76 | 97.52 | 96.88 | 99.91 | 98.68 | 99.99 |
| BFC | 98.40 | **100.00** | 98.23 | **100.00** | 97.41 | 99.98 | 89.33 | 98.14 | 96.65 | **99.98** | 98.39 | **100.00** |
| BLESS | 99.83 | **100.00** | 99.85 | 99.99 | **99.23** | 99.98 | **92.80** | 99.08 | **98.59** | **99.98** | 98.65 | **100.00** |
| Blue | 99.64 | 99.90 | 99.68 | 99.92 | 96.07 | 99.66 | 84.35 | 92.67 | 93.08 | 99.73 | **98.78** | 99.94 |
| Coral | 92.13 | 98.72 | 92.52 | 98.52 | 79.26 | 97.84 | 51.26 | **99.54** | N/A | N/A | 95.96 | 99.57 |
| ECHO | 87.46 | 99.99 | 93.33 | 99.99 | 88.52 | 99.94 | N/A | N/A | N/A | N/A | 94.98 | **100.00** |
| Fiona | 98.28 | 99.96 | 98.65 | 99.94 | 95.28 | 99.76 | 70.46 | 94.31 | N/A | N/A | 98.17 | 99.99 |
| HiTEC | 98.78 | 99.99 | 99.30 | 99.99 | N/A | N/A | N/A | N/A | N/A | N/A | 97.83 | **100.00** |
| Lighter | 99.30 | **100.00** | 99.47 | **100.00** | 98.13 | **99.99** | 79.71 | 99.33 | 96.13 | **99.98** | 98.22 | **100.00** |
| Musket | 99.49 | **100.00** | 99.50 | **100.00** | 97.87 | 99.98 | 84.32 | 98.33 | 93.86 | **99.98** | 97.79 | **100.00** |
| Quake | 99.57 | **100.00** | 99.58 | **100.00** | 98.41 | **99.99** | 88.17 | 98.76 | 94.71 | **99.98** | 95.82 | **100.00** |
| QuorUM | **99.88** | 100.00 | **99.90** | 100.00 | 98.78 | 99.98 | 86.54 | 98.74 | 97.29 | **99.98** | 98.64 | 99.99 |
| RACER | 98.51 | 99.96 | 99.29 | 99.96 | 96.40 | 99.94 | 74.16 | 99.24 | 92.95 | 99.95 | 98.36 | 99.99 |
| Reptile | 97.77 | 99.99 | 98.25 | 99.97 | 92.00 | 99.86 | 79.47 | 97.22 | 89.65 | 99.92 | 96.69 | 99.99 |
| SGA | 99.57 | **100.00** | 99.60 | **100.00** | 98.53 | **99.99** | 86.72 | 98.87 | 97.61 | **99.98** | 97.95 | **100.00** |

[Aligned] the percentage of aligned reads to the total number of reads; [Correct]: the ratio of reads that were aligned to correct positions as a percentage of the total number of aligned reads; Original: pre-correction results.

In almost all the cases, the ratio of correctly aligned reads to the total number of aligned reads was over 99 percent with the exception of I4. For I4, only the corrected reads from BLESS, Lighter, and Racer showed the accuracy of over 99 percent.

**Runtime and Memory Usage of Illumina Error Correction Tools**

To compare how the run time and memory usage of various tools scale with the size of the input, we compared each Illumina error-correction method for two cases, I5-20X and I5-40X which has twice the number of reads as I5-20X. These results are summarized in Table 8. Except Reptile, all the evaluated Illumina error-correction tools support parallelization, and 12 threads were used for the tools. In addition to running parallel threads on a single node, BLESS can also be parallelized across multiple nodes using MPI. BLESS's results on two computing nodes are reported separately. For I5-40X, BLESS, Lighter, and SGA could correct the read set using under 4 GB of memory. BFC, BLESS, Blue, Lighter, QuorUM, and RACER used almost the same memory for both 20 X and 40 X coverage reads. The fastest tools were BLESS and Lighter and they were over 13 times faster than ALLPATHS-LG. ALLPATHS-LG required 3.6 times longer time for correcting I5-40X than I5-20X.

Table 8. Memory usage and runtime of Illumina error correction tools for I5-20X and I5-40X.

| Software | Memory Usage (MB) | | Runtime (min) | |
| --- | --- | --- | --- | --- |
| | I5-20X | I5-40X | I5-20X | I5-40X |
| ALLPATHS-LG | 12,287 | 18,424 | 122 | 435 |
| BFC | 10,753 | 10,889 | 12 | 21 |
| BLESS (1 node) | 3,813 | 3,825 | 9 | 15 |
| BLESS (2 nodes) | 3,809 | 3,799 | **5** | **9** |
| Blue | 20,286 | 20,398 | 29 | 46 |
| Lighter | **1,107** | **1,109** | 9 | 13 |
| Musket | 4,215 | 6,647 | 19 | 36 |
| Quake | 13,760 | 21,643 | 74 | 143 |
| QuorUM | 8,163 | 8,686 | 10 | 22 |
| RACER | 12,623 | 14,490 | 17 | 35 |
| Reptile | 13,016 | 17,422 | 815 | 1,711 |
| SGA | 1,874 | 3,508 | 61 | 125 |
| SOAPec | 4,985 | 9,708 | 42 | 71 |

**Effect of Using Different Alignment Tools on the Evaluation of Real Reads**

For real reads, we compare the errors corrected by an error-correction tool against mismatches and indels obtained in aligning the reads to a reference sequence. Therefore, the numbers and the locations of errors could vary according to alignment tools. We generated two $F_L$ files from I6 using BWA [43] and Bowtie 2 [51] with default options, and the two files were compared. While BWA found 473,090 substitution errors in D6, Bowtie 2 found 632,705. About 97 percent of substitutions in the BWA set were also found in the Bowtie 2 set, which means Bowtie 2 is more aggressive than BWA and it could indicate more errors in reads.

When the error-correction results were evaluated using the $F_L$ file from Bowtie 2, sensitivity and gain dropped by up to 8 percent compared to the results with the $F_L$ file from BWA because some of the new errors found by Bowtie 2 were not corrected in the error correction tools. Detailed results can be found in the supplementary document.

## EVALUATION RESULTS FOR TGS ERROR-CORRECTION TOOLS

Due to the high error rate of TGS reads, error correction outputs could have many uncorrected bases. Therefore, most TGS error-correction tools generate two types of reads: 1) trimmed reads that only contain corrected regions in input reads and 2) untrimmed reads that include both corrected and uncorrected regions in input reads.

For PacBio reads, PBcR only produces trimmed reads, LSC and Proovread generate both trimmed reads and untrimmed reads, and they were assessed separately. For LoRDEC, trimmed reads were generated from the untrimmed reads using lordec-trim-split that is included in the LoRDEC package. For MinION reads, both NanoCorr and NaS produce trimmed reads.

### Accuracy of PacBio Error Correction Tools

In Figure 5A, percentage similarity of the outputs from PacBio read error-correction methods for P1 are compared. Percent similarity of the input reads was 76.6% before error correction, and all the output results were better than this number. Among the four tools, three tools except LSC showed percentage similarity over 95% for the trimmed reads. For the untrimmed reads, LoRDEC and Proovread generated more accurate reads than LSC. Except for the case of untrimmed LoRDEC reads, read coverage of Illumina reads had almost no impact on percentage similarity.

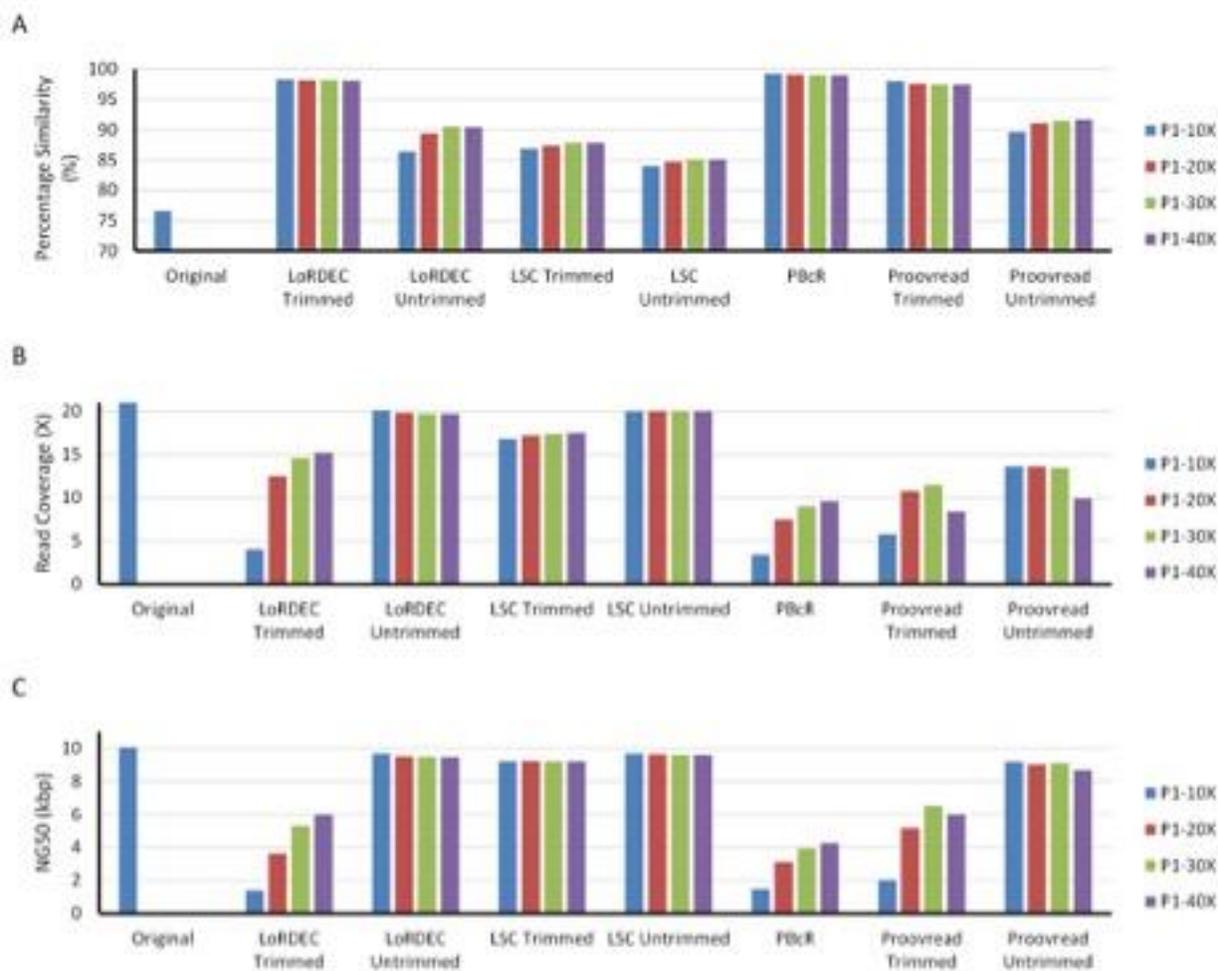

**Figure 5.** Percentage similarity, read coverage, and NG50 of PacBio read error correction methods for P1. [Original] results for the input PacBio reads before error correction.

Figure 5B and Figure 5C show read coverage and NG50 of the outputs of the compared tools. The two charts had similar shapes. Both values were high where percentage similarity in Figure 5A was low. The trimmed LoRDEC reads and the PBcR outputs were improved a lot by increasing Illumina read coverage. The trimmed reads from Proovread were also improved but the values were saturated at 30 X coverage.

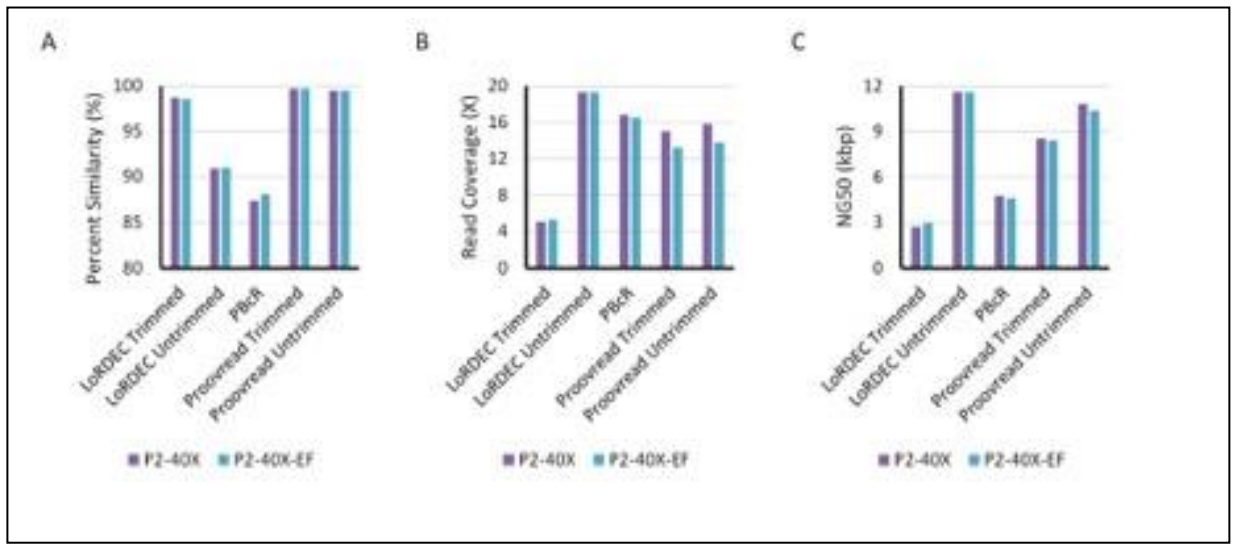

**Figure 6.** Percentage similarity, read coverage, and NG50 of PacBio read error correction methods for P2. Percentage similarity, read coverage, and NG50 of the input PacBio reads before error correction were 79.4 percent, 20 X, and 12,095 bp, respectively.

Percentage similarity, read coverage, and NG50 are compared for P2-40X and P2-40X-EF that is the error-free version of P2-40X in Figure 6. Both the trimmed Proovread reads and the trimmed LoRDEC reads showed high percentage similarity. Percentage similarity and read coverage of the untrimmed Proovread reads were almost the same compared to those of the trimmed Proovread reads. However, NG50 of the trimmed Proovread reads was shorter than that of the untrimmed Proovread reads. LoRDEC generated the trimmed reads with high percent similarity but it removed too many bases and read coverage and NG50 of the read set became much lower than those of the original input reads.

For all these cases, P2-40-EF did not make a meaningful difference when it was compared with P2-40. This means sequencing errors in Illumina reads are not important when Illumina read coverage is about 40 X.

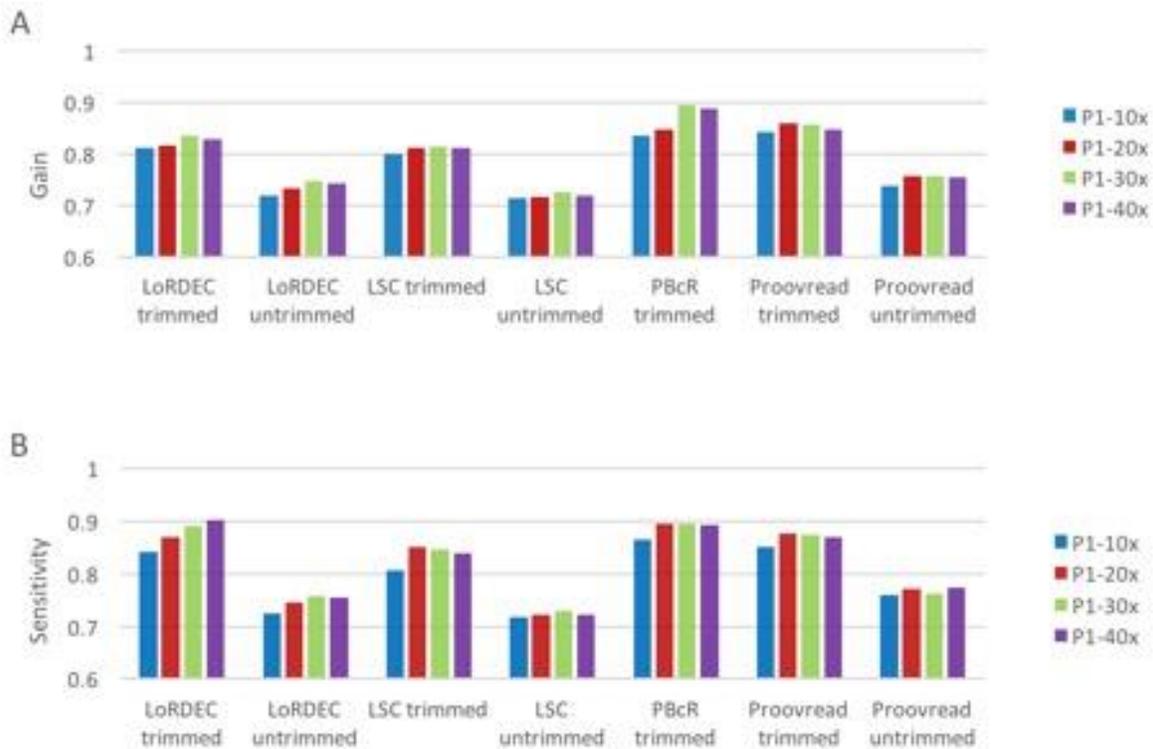

**Figure 7.** Sensitivity and gain of Pacbio reads.

Figure 7A and 7B show the sensitivity and gain results for the different PacBio error-correction tools. These results include both indel and substitution errors. Compared to tools correcting Illumina reads, the PacBio error-correction tools seem to have lower sensitivity and gain. Similar to percentage similarity for the same dataset, gain and sensitivity generally improve upon trimming. For example, it may be seen that the sensitivity of trimmed reads of LORDEC, Proovread and LSC are significantly higher than that of untrimmed versions. For LORDEC trimmed reads, though sensitivity increases with higher Illumina coverage, gain remains largely unchanged indicating that at higher Illumina coverage, more errors are corrected, but also there are more false corrections.

**Accuracy of ONT Read Error Correction Tools**

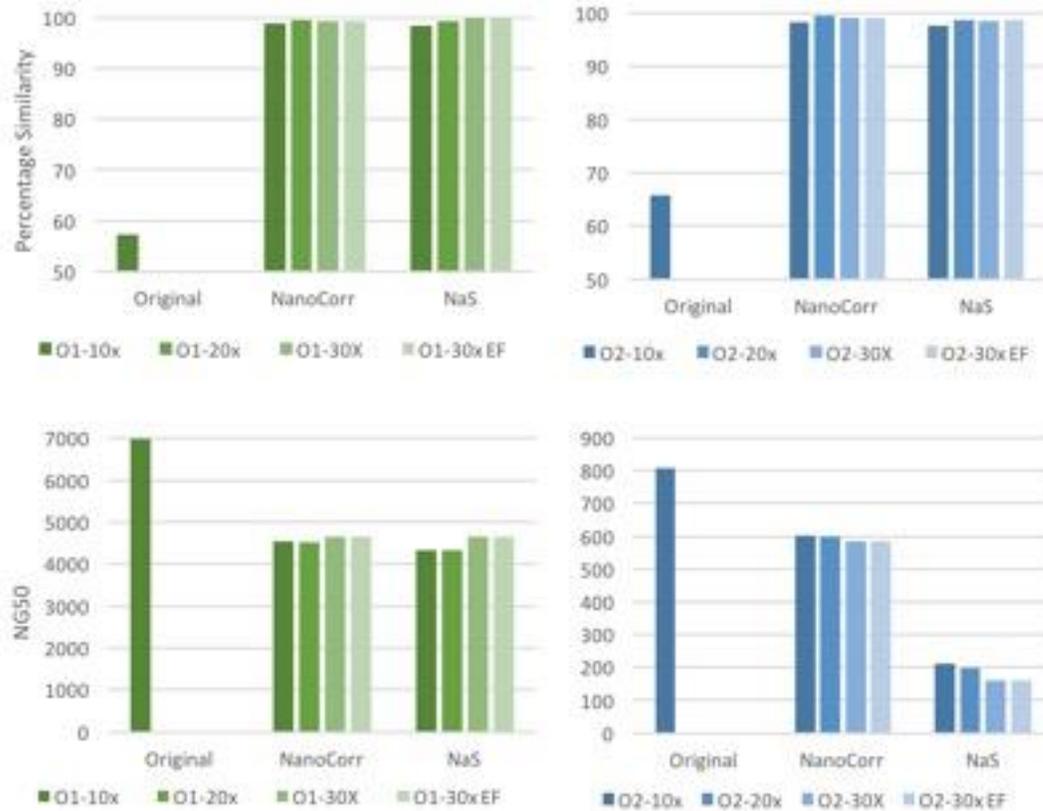

**Figure 8.** Percentage similarity and NG50 lengths of input MinION reads.

Figure 8A shows percentage read similarity values for ONT datasets. For O1, the original read similarity was 57.3%, which is lower compared to the corresponding PacBio reads (comparing P1 and O1 from the E.Coli genome). Both the error-correction tools significantly improved the percentage similarity of reads, and the values did not significantly change with different coverage values of the corresponding Illumina datasets. Figure 8B shows the NG50 values for O1 and O2 datasets. Both tool outputs have a lower NG50 length than the input reads and NaS reads have a noticeably lower NG50 length compared to NanoCorr for the O2 dataset. Using error-free Illumina reads did not bring in a noticeable improvement in error correction. Figures 9A and 9B summarize the sensitivity and gain for the two ONT datasets. These results include both indel and substitution errors. It may be noted that NaS presents slightly higher sensitivity and gain compared to NanoCorr in both cases.

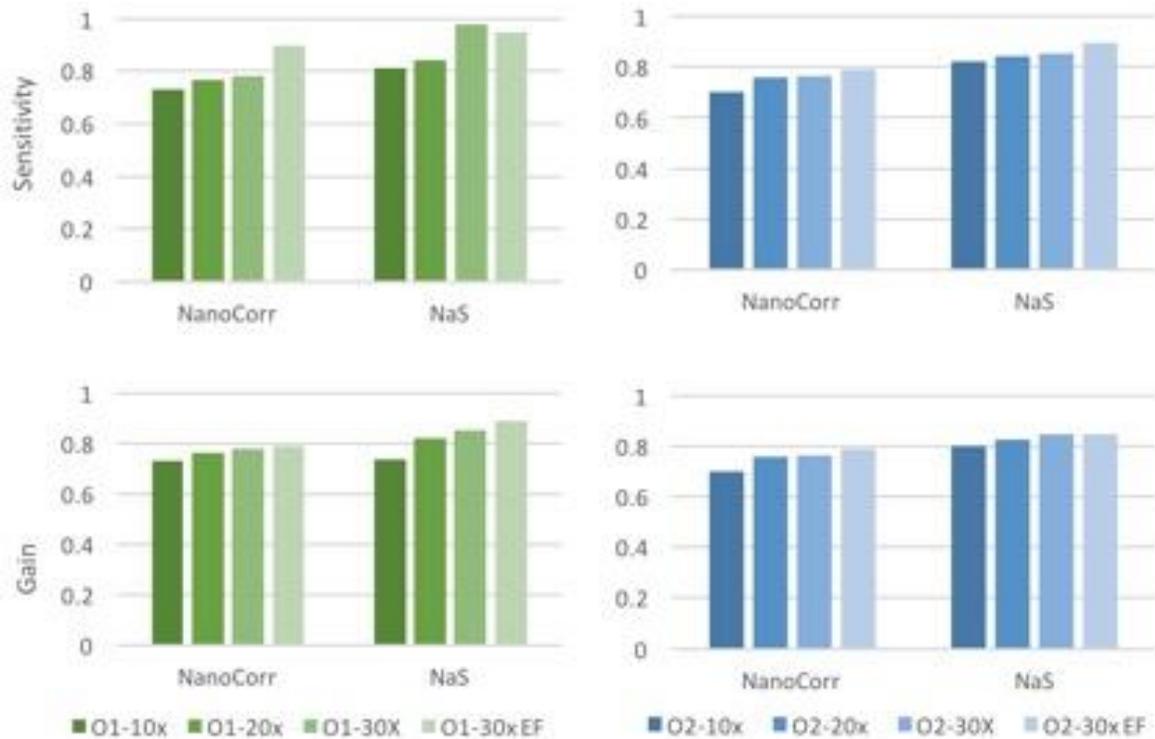

**Figure 9.** Sensitivity and gain of MinION reads.

## CONCLUSIONS AND DISCUSSION

Among the Illumina read error correction methods that were evaluated, ALLPATHS-LG, BFC, BLESS, Lighter, Quake, QuorUM, and SGA generated accurate results for over 30 X read coverage. BLESS and Quake outperformed the others for reads with 10-20 X read coverage, and it is expected that ALLPATHS-LG would work best for the reads with under 10 X read coverage when we see Figure 3. For highly repetitive genomes, it is recommended to use BLESS and Quake for getting the most accurate results.

Among the evaluated PacBio error correction tools, there was no apparent winner that could generate both accurate and long reads. While trimming improved the accuracy of correction significantly, it reduced the NG50 length and read coverage appreciably. Proovread may be recommended in cases where the accuracy of corrected reads is more important than their length. If a large read set should be corrected in a short time, LoRDEC might be a good choice.

Both tools evaluated for ONT reads provided outputs with good percentage similarity and had comparable gain and sensitivity with respect to the PacBio tools. NanoCorr had longer NG50 length for one of the datasets. ONT sequencing techniques have a higher error rate compared to PacBio sequencing techniques. However, the evaluated tools seem to improve the accuracy of ONT reads significantly. Given that ONT sequencing is cheaper, easy to use and provides higher data throughput, further improvements to existing error-correction methods for ONT reads as well as developing read simulation techniques to support the evaluation of these tools can make it a very attractive alternative.

Though some tools have recommendations for choosing input parameters, we tried to tune the parameters independently based on the results for fair comparison. However, in a real situation where the locations of errors are not known in advance, it would not be possible to find the best parameters this way. Therefore, it is recommended to developers of error-correction tools that as many parameters as possible should be automatically determined or clear guidelines for determining them should be given to users.

We believe that SPECTACLE will also be compatible with new sequencing technologies and some of its potential is evident from the fact that it can work with NGS and TGS reads with varied characteristics, providing a comprehensive set of evaluation metrics. The fundamental strength of the tool is that the underlying evaluation algorithms are not tied to specific read lengths or error models.

Even though the work presents a comprehensive analysis of the accuracy of most of the state-of-the-art error correction methods for the NGS and TGS technologies, the study can be further extended to evaluate TGS reads from larger genomes using more powerful computational resources. It is expected that repeats in a genome would affect the quality of error correction in TGS reads. However, repeat effects will play a significant role in correction only when genome length is sufficiently long. We tried using a PacBio sequence of the 10 Mbp regions of mouse chromosome Y but their results were not worse than those collected for P2 corrected by LoRDEC and Proovread. These results were not included in the main text because the genome was too short for us to reach useful conclusion regarding repeats. The results are in the supplementary document.

It is also desirable to study how sequencing errors degrade the quality of downstream analysis pipelines like variant calling. A detailed understanding of the mechanism will yield useful insights into how to correct errors that are detrimental to a specific application and how to make applications less sensitive to sequencing errors, and SPECTACLE can help in categorizing such errors.

## SOFTWARE AVAILABILITY

SPECTACLE and all the input reads and run scripts used in these experiments are available at https://github.com/gowthami19m/SPECTACLE.

## SUPPLEMENTARY DATA

Supplementary data are available online at https://github.com/gowthami19m/SPECTACLE.

## KEY POINTS

- Correcting errors in high throughput sequencing data can improve the quality of downstream analyses.
- We developed a software package that can evaluate the error correction tools for DNA/RNA reads and NGS/TGS reads, and compared 23 state-of-the-art error correction methods.
- Recommended tools for Illumina reads are ALLPATHS-LG, BFC, BLESS, Lighter, Quake, QuorUM, and SGA in general, BLESS and Quake for repetitive genomes, and ALLPATHS-LG for reads with under 10 X coverage.
- Recommended tools for PacBio reads are Proovread for accurate error correction and LoRDEC for scalability. There is a significant trade-off between output read-length and output read accuracy in the case of PacBio error correction.
- Though ONT sequencing reads have a higher error rate than PacBio reads, the two ONT tools improve the accuracy significantly and provide output reads with quality comparable to PacBio error corrected reads.

# BIOGRAPHY

Yun Heo completed his PhD at the University of Illinois at Urbana-Champaign from the department of Electrical and Computer Engineering.

Gowthami Manikandan is a Master's student in the department of Electrical and Computer Engineering at the University of Illinois at Urbana-Champaign.

Anand Ramachandran is a PhD student in the department of Electrical and Computer Engineering at the University of Illinois at Urbana-Champaign.

Deming Chen is a Professor in the department of Electrical and Computer Engineering at the University of Illinois at Urbana-Champaign.